\documentclass[lettersize,journal]{IEEEtran}
\usepackage{amsmath,amsfonts}
\usepackage{algorithmic}
\usepackage{algorithm}
\usepackage{array}
\usepackage[caption=false,font=footnotesize,labelfont=bf,textfont=bf]{subfig} 
\usepackage{textcomp}
\usepackage{stfloats}
\usepackage{url}
\usepackage{verbatim}
\usepackage{graphicx}
\usepackage{xcolor}
\usepackage{cite}
\usepackage{enumerate}
\usepackage{epstopdf}
\usepackage{booktabs}
\usepackage{makecell}
\usepackage{float}
\usepackage{multirow}
\usepackage{tabularx}
\usepackage{booktabs}    
\usepackage{array}       
\usepackage{makecell}    

\begin{document}

\title{Simultaneous Self-Localization and Base Station Localization with Resonant Beam}

\author{}
\author{Guangkun Zhang, Wen Fang, Mingliang Xiong, Qingwen Liu,~\IEEEmembership{Senior Member,~IEEE}, Mengyuan Xu,\\
Yunfeng Bai,~\IEEEmembership{Member,~IEEE}, Mingqing Liu, Siyuan Du
\thanks{Guangkun Zhang and Qingwen Liu are with Shanghai Research Institute for Autonomous Intelligent Systems, Tongji University, Shanghai 201804, China. (e-mail: 2210993@tongji.edu.cn; qliu@tongji.edu.cn)}
\thanks{Wen Fang, Mengyuan Xu, Yunfeng Bai, Mingqing Liu, and Siyuan Du are with the College of Electronics and Information Engineering, Tongji University, Shanghai 201804, China. (e-mail: wen.fang@tongji.edu.cn; xumy@tongji.edu.cn; baiyf@tongji.edu.cn; ml2176@cam.ac.uk; dueen1123@tongji.edu.cn)}
\thanks{Mingliang Xiong are with the School of Computer Science and Technology, Tongji University, Shanghai 201804, China. (e-mail: mlx@tongji.edu.cn;) }
}

\maketitle

\begin{abstract}
High-precision positioning in GPS-denied environments is a challenging yet critical technology. Resonant beam positioning (RBP) utilizes a resonant beam with properties such as energy focusing, self-establishment, self-alignment, and passive operation, offering a promising solution for this task. However, traditional RBP algorithms require a fixed number of resonant beam base stations, which can be costly to expand coverage. To address this limitation, we propose a distributed resonant beam positioning (DRBP) system that simultaneously estimates the base station and mobile target (MT) positions. Firstly, the MT receives resonant beam samples to locate the base station in the limited field of view (FoV) region. Subsequently, it estimates its own position by tracking the apparent motion of these newly localized base stations. During moving, the DRBP system facilitates self-positioning on the MT side, enabling dynamic expansion of both the number of base stations and the coverage area. Numerical results demonstrate that DRBP achieves a positioning root mean square error (RMSE) of $0.1$ m and a rotation RMSE of 2$^\circ$, validating the system's high accuracy.
\end{abstract}


\section{Introduction}

In the internet of everything (IoE) era, applications such as intelligent unmanned systems, smartphones, and virtual reality increasingly demand high-precision positioning \cite{ref1, ref2, ref3}. GPS is the most widely adopted technology for positioning, offering extensive global coverage and reliable accuracy \cite{ref8}. However, signal blockages and reflections often compromise GPS performance, particularly in urban environments or areas with limited satellite visibility. In scenarios where GPS is unavailable, various positioning techniques have been extensively explored, including UWB, motion capture, visual simultaneous localization and mapping (SLAM) and lidar SLAM. Ultra-wideband (UWB) technology achieves sub-meter-level positioning accuracy (down to 30 cm) through distributed downlink time difference of arrival schemes, enabling tracking any number of assets without decreasing the measurement update rate \cite{ref10, ref11}. In contrast to UWB technology, motion capture systems such as vicon achieve centimeter-level positioning accuracy by leveraging multi-camera infrared setups \cite{ref39, ref40, ref41}. However, vicon systems have high calibration and expensive costs, limiting their practicality in specific applications. Visual SLAM and lidar SLAM has minimal reliance on an external base station and can achieve centimetre-level accuracy in controlled datasets within short time frames \cite{ref14,ref15,ref16}. However, SLAM has limitations: Visual SLAM struggles in low-light conditions, and LiDAR SLAM may face difficulties in feature-poor or repetitive environments, such as long corridors \cite{ref17}.

\begin{figure}[!t]
\centering
\includegraphics[width=3in]{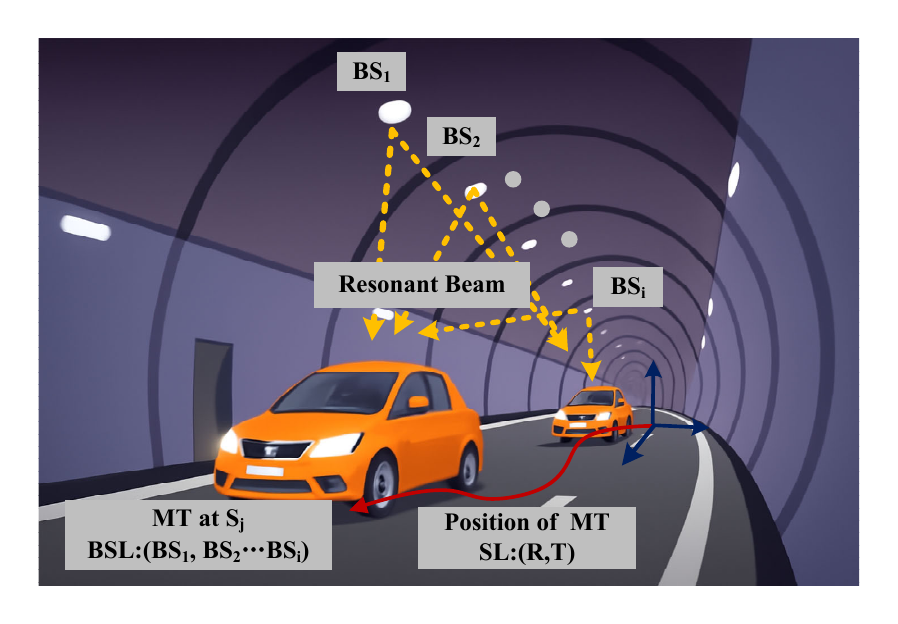}
\caption{A typical application scenario of the DRBP. Mobile target (MT) at state i (S$_j$) using base station localization (BSL) method to estimate unknown base stations' positions in MT coordinate system. The self-localization (SL) method calculates the positional changes of base stations within the field of view in the MT coordinate system to estimate MT rotation and displacement, i.e., MT position.}
\label{fig_1_1}
\end{figure}

The resonant beam system (RBS) is an advanced technology for over-the-air power transfer \cite{ref22, ref23}, wireless communication \cite{ref24,ref25}, and wireless positioning \cite{ref26,ref27}. A standard RBS configuration consists of two retroreflectors and a gain medium, which forms a resonator. The gain medium, combined with one retroreflector, forms the transmitter (RB-T), while the other retroreflector acts as the receiver (RB-R). If RB-R is within RB-T's field of view (FoV), the wireless link between RB-T and RB-R can be automatically established and self-aligned without the need for RB-R to transmit a signal actively \cite{ref31, ref32}. Furthermore, due to the feedback amplification structure of the resonant cavity, the beam energy is retained and amplified within the cavity, resulting in a high signal energy concentration. With energy focusing, self-establishing, and passive characteristics, RBP systems present a solution for high precision positioning technologies in GPS-denied environments \cite{ref26}. A typical RBP system implementation employs multiple RB-Ts, utilizing binocular disparity for depth estimation and centroid-based angle of arrival (AoA) calculation, achieving within 5 cm accuracy in a 2-meter range \cite{ref30}. {\color{blue} A comparison of major positioning techniques is summarized in Table I}. We found that the RBP system based on known base station positions has the advantages of high precision, passive and self-alignment, and self-establishment. However, existing RBP systems have only been evaluated with a fixed number of base stations. Thus, the retroreflector's optical characteristics limit the effective positioning range, resulting in limited coverage, i.e., the FoV of RB-T \cite{ref34}. Integrating multiple systems to expand coverage inevitably encounters communication and mobile target (MT) identification challenges.

{\color{blue} This paper presents a novel distributed resonance beam positioning (DRBP) system that achieves simultaneous self-localization and base station localization, as highlighted in Table I. In this system, each MT is equipped with a wavefront sensor to measure the AoA of the resonant beam, while the base stations are low-cost, passive units retrofitted with retroreflectors. By combining AoA measurements with prior knowledge of the base station deployment pattern, an MT can estimate the positions of the base stations within its FoV. As the MT moves, it tracks the positional shifts of these base stations to compute its own displacement in distance and direction, thereby enabling continuous and accurate self-localization. This MT-centric architecture is inherently designed for high concurrency and scalability, as illustrated in Fig.~\ref{fig_1_1}. Since each MT performs all computations locally and the base stations operate entirely passively, multiple MTs can operate independently within the same area, sharing the common set of base stations without contention or a central coordination bottleneck. Consequently, the DRBP framework effectively overcomes the limitations in base station extensibility and multi-MT concurrency that are inherent in traditional RBP systems.} The main contributions of this work are as follows:

\begin{enumerate}
\item We propose a distributed resonance beam positioning (DRBP) system, which removes the need for a fixed number of base stations and resolves the multi-MT concurrency challenges inherent in centralized systems. DRBP employs base station localization (BSL) to estimate the base station position and utilizes self-localization (SL) to estimate the MT position, enabling the extension of base station quantities and eliminating the constraints of positioning algorithms on multi-MT concurrency. Leveraging resonant beam self-established, self-alignment, and energy focusing character, DRBP achieves high-precision positioning.
\item We develop a theoretical model for evaluating DRBP performance. By accurately simulating the propagation of resonant beams within the transmission channel, we estimate the AoA error range and the precision of BSL and SL, guiding system performance evaluation and optimization. Numerical results show that BSL achieves sub-decimeter positioning accuracy with a root mean square error (RMSE) of $0.18$ m, even with AoA errors up to $0.2^{\circ}$. The SL accuracy achieves a positioning RMSE of $0.1$ m and a rotational RMSE of $2^{\circ}$.
\end{enumerate}

The remainder of this article is organized as follows. Section~II outlines the system architecture and fundamental principles of the DRBP. Section III details the BSL methods for AoA estimation using resonant beam phases and wavefront sensors, along with the depth calculation process through the RB-T deployment mode and disparity principle. Section IV presents the principles of SL via resonant beam, including the installation, coordinate transformations, minimal system deployment, and the method for optimizing position accuracy in multi-base station scenarios. Additionally, system errors related to beam power cycles and noise factors are analyzed. Section V presents simulation results validating the system's effectiveness. Finally, Section VI concludes the study.

\section{System Overview}
\label{sec:so}
\begin{figure}[!t]
\centering
\includegraphics[width=3.5in]{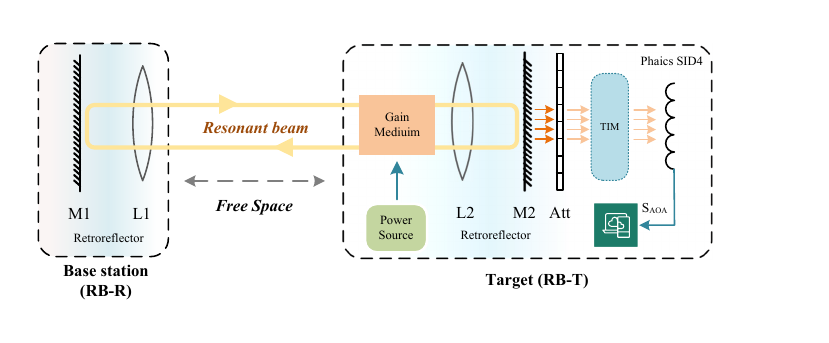}
\caption{System diagram (TIM: telescope internal modulator; $S_{AoA}$: optical field distribution on wavefront sensor); Att: attenuator.}
\label{fig_1}
\end{figure}

\begin{table*}[htb]
\caption{{\color{blue}Comparison of Positioning Systems}}
\label{table_compare}
\centering
\renewcommand{\arraystretch}{1.3} 
\begin{tabular}{
    >{\raggedright\arraybackslash}m{4cm} |
    >{\raggedright\arraybackslash}m{3cm} |
    >{\raggedright\arraybackslash}m{9cm}
}
\hline
\textbf{System} & \textbf{Accuracy} & \textbf{Features} \\
\hline
GPS \cite{ref8} & Meter-level & - Easy to deploy with no cumulative errors. \newline - Sensitive to environmental factors and signal interference. \\
\hline
UWB \cite{ref10, ref11} & Submeter-level & - High-precision localization within 15 meters down to 30 cm. \newline - Sensitive to electromagnetic interference. \\
\hline
Vision/Lidar SLAM \cite{ref14, ref15, ref16, ref17} & Centimeter-level & - Flexible deployment without predefined environments. \newline - Sensitive to lighting and electromagnetic interference, with high computational demands. \\
\hline
Vicon \cite{ref39, ref40, ref41, ref42} & Sub-millimeter & - Extremely high accuracy with infrared cameras tracking reflective markers. \newline - Requires complex camera setup, precise calibration, and is costly. \\
\hline
RBP \cite{ref22, ref23, ref24, ref25, ref26, ref27} & Centimeter-level & - No need for manual calibration, passive and simplifying deployment. \newline - Energy-focused fature with no signal scattering. \\
\hline
Traditional RBP \cite{ref26, ref27} & Centimeter-level & - High accuracy with energy-focused beam. \newline - Fixed number of base stations, limited scalability. \newline - Centralized architecture, potential concurrency issues. \\
\hline
\textbf{Proposed DRBP} & \textbf{Centimeter-level} & \textbf{- All advantages of RBP.} \newline \textbf{- Distributed architecture, high scalability and concurrency.} \newline \textbf{- Low-cost, passive base stations simplify deployment.} \newline \textbf{- Simultaneous self-localization and BS-localization.} \\
\hline
\end{tabular}
\end{table*}

\subsection{System Design}
{\color{blue}The proposed DRBP system establishes a positioning framework where the reference coordinate system is defined by the MT. Two coordinate systems are utilised: the initial coordinate system, which is fixed to the MT's state at an arbitrary starting moment ($t=0$), and the MT coordinate system, which is a local frame that moves with the MT. The objective of the DRBP system is to determine the MT's pose (position and orientation) within the initial coordinate frame at any given time $t$. This MT-centric approach creates a distributed architecture, decoupling the positioning algorithm from any fixed base station infrastructure and thereby enhancing system scalability and concurrency.

As illustrated in Fig.~\ref{fig_1}, the DRBP consists of two main components: RB-R and RB-T. The RB-T is installed on the MT for positioning, while a single RB-R is passively deployed on the ground as a base station. The RB-R only include a retroreflector, and the RB-T is equipped with a retroreflector, a gain medium, an attenuator (Att), and a wavefront sensor. This wavefront sensor measures the beam distribution, combining with the MUSIC algorithm to estimate AoA \cite{ref34}. The RB-T also includes a telescope internal modulator (TIM) for phase correction \cite{ref32}. An RB-T and an RB-R form a spatially distributed laser resonator within which a resonant beam is generated.

The resonant beam exhibits self-establishing and self-aligning properties. Specifically, the gain medium at the RB-T end amplifies omnidirectional incident light, generating a low-energy field covering its entire FoV. When the RB-R enters this FoV, its retroreflector reflects a portion of the amplified low-energy light directly back to the RB-T. This reflection triggers a feedback loop between the RB-T and RB-R, ultimately leading to the constructive interference of in-phase waves and the iterative cancellation of out-of-phase components, thereby forming a resonant beam \cite{ref28}. Since the feedback propagates at the speed of light, the energy is rapidly amplified, resulting in the formation of a well-collimated, precisely aligned resonant beam with concentrated energy. The resonant beam is formed naturally when the spatially distributed laser resonator remains FoV even though the RB-R is moving \cite{ref45, ref34}.

\subsection{Process of Positioning}
DRBP provides continuous position services for MTs. Since the coordinate system is solely determined by the MTs, the positioning service must account for the MT's rotation. As the MT moves, the apparent positions of the static base stations change within the MT coordinate system. By tracking these changes, the system accurately estimates the MT's displacement and rotation relative to the initial coordinate system. The positioning methodology unfolds in three sequential stages:

\begin{enumerate}
\item AoA Measurement: At each time step, the RB-T's wavefront sensor captures the light field distribution of the resonant beam from each visible base station. The MUSIC algorithm is then applied to this data to precisely estimate the unit direction vector, $\mathbf{x}_{i}$, corresponding to the AoA of the beam from the $i$-th base station (Section \ref{sec:ac}).

\item BSL: Leveraging the known deployment pattern of the RB-Rs, the system calculates the distance (depth) $d_{i}$ to each base station from the MT (Section \ref{sec:dc}). By combining the estimated AoA vectors ($\mathbf{x}_{i}$) the precise 3D coordinates of each visible base station are determined within the current MT coordinate system.

\item SL: With the positions of the base stations localized in the MT coordinate system, the final stage estimates the MT's own motion. By comparing the set of base station coordinates between consecutive states, the algorithm computes the incremental rotation matrix $\Delta \mathbf{R}$ and translation vector $\Delta \mathbf{T}$ that describe the MT's movement (Section \ref{sec:pc}). This step solves for the MT's motion, which is then used to update its absolute pose in the initial coordinate system.
\end{enumerate}

\subsection{System Initialization and Recursive Process}
The initialization procedure of the DRBP algorithm corresponds to the process of defining the initial coordinate system. This process is triggered only when the minimum number of base stations required by the DRBP system appears within the FoV of the MT. The initialization time is designated as \(t = 0\), at which the pose of the MT is set to the origin with an identity orientation. Mathematically, this is represented by the rotation matrix \(\mathbf{R}_0 = \mathbf{I}\) and the translation vector \(\mathbf{T}_0 = \mathbf{0}\).

The characteristic of this initialization is its rapidity. The computational load of the AoA estimation algorithm dominates the total time overhead. At the same time, the physical establishment of the resonant beam link, governed by speed-of-light dynamics, is effectively instantaneous and thus considered negligible in comparison. The entire initialization procedure is typically completed in approximately 0.1 seconds, enabling near-instantaneous operation once the geometric conditions are met.

After initialization, the MT recursively estimates its incremental pose change \((\Delta \mathbf{R}_t, \Delta \mathbf{T}_t)\) over each time interval \([t, t+1]\) through iterative AoA measurement, BSL, and SL. A key aspect of this process is the dynamic management of the set of visible base stations \(\mathbf{P}_t\), which involves data association between consecutive frames. Specifically, a nearest-neighbour search with a distance threshold is used to match currently observed base stations with those from the previous time step. Observations that do not match any existing base station within the threshold are registered as new base stations, while previously tracked base stations that are no longer observed are removed from the set.

The absolute pose of the MT in the initial coordinate system, denoted \((\mathbf{R}_{t+1}, \mathbf{T}_{t+1})\), is updated by composing the previous pose \((\mathbf{R}_t, \mathbf{T}_t)\) with the incremental transformation. The updated rotation is given by:
\begin{equation}
\mathbf{R}_{t+1} = \mathbf{R}_t \cdot \Delta \mathbf{R}_t.
\end{equation}
The updated translation is computed as:
\begin{equation}
\mathbf{T}_{t+1} = \mathbf{T}_t + \mathbf{R}_t \cdot \Delta \mathbf{T}_t.
\end{equation}
Through this recursive mechanism, the DRBP algorithm accumulates incremental motions, allowing the MT to accurately track its absolute position and orientation relative to the initial coordinate system at all times.
}

\section{Base Station Positioning Method}
\label{sec:pe}

In the DRBP, it is essential to obtain the positions of the base station within the MT coordinate system. This section introduces BSL for positioning the base station on the MT side. We first introduce the method for measuring the incident angle of the resonant beam on RB-T. Next, we incorporate the base station deployment pattern and detail BSL. Finally, we discuss the positioning approach in the presence of individual base station occlusions to enhance the DRBP robustness.

{\color{blue}
\subsection{Angle Estimation}
\label{sec:ac}

Accurate AoA measurements are the foundation of the BSL positioning process. The AoA is estimated from the spatial distribution of the resonant beam's optical field as captured by a wavefront sensor, as shown in Fig.~\ref{fig_aoa}. Therefore, a comprehensive understanding of the AoA estimation process requires two key components: first, a physical model of the resonant beam to characterise the signal source, and second, the signal processing algorithm used to extract the directional information. Accordingly, this subsection first presents the theoretical models for the resonant beam under both static and dynamic conditions, and then details the application of the MUSIC algorithm for AoA estimation.

\begin{figure}[!t]
\centering
\includegraphics[width=3in]{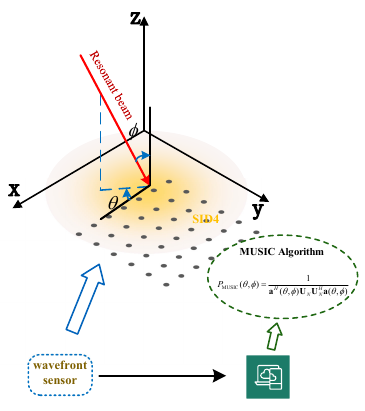}
\caption{Schematic of the AoA estimation process, depicting the distribution of elevation and azimuth angles on the wavefront sensor, and the resolution of the optical signal using the MUSIC algorithm.}
\label{fig_aoa}
\end{figure}

\subsubsection{Static Cavity Model and Field Distribution}
When a base station is within the MT's FoV, a stable resonant beam is formed. This beam is characterized by a self-reproducing eigenmode, which is the steady-state field distribution within the optical resonant cavity. To determine this eigenmode, we employ the classic Fox-Li numerical iteration method \cite{ref26}. Within the fourier optics framework, the propagation of the beam through various intracavity elements such as mirrors ($\Gamma_{\text{M}}$), lenses ($\Gamma_{\text{L}}$), gain media ($\Gamma_{\text{G}}$), and free space ($\Gamma_{\text{FS}}$) can be described by corresponding mathematical operators.

The self-reproducing mode, denoted by $u(x, y)$, is the eigenfunction of the round-trip propagation operator $\Gamma_{\text{round}}$. This relationship is expressed by the fundamental cavity equation:
\begin{equation}
\Gamma_{\text{round}}(u) = \gamma u,
\end{equation}
where $\gamma$ is the transmission factor. During the iterative numerical solution process (e.g., the Fox-Li method), where $u_{t+1} = \Gamma_{\text{round}}(u_t)$, the transmission factor is estimated at each step $t$ by comparing the total amplitude of the output field to that of the input field:
\begin{equation}
\gamma_t = \frac{\iint u_{t+1}(x, y) \, dx \, dy}{\iint u_t(x, y) \, dx \, dy}.
\end{equation}
As the iteration converges, $\gamma_t$ stabilizes to the final value $\gamma$. The magnitude of this converged factor, $|\gamma|$, determines the round-trip power efficiency, given by $|\gamma|^2$, and the corresponding power loss is $1 - |\gamma|^2$.

The operator $\Gamma_{\text{round}}$ concatenates all optical transformations in a single round trip. For a typical linear cavity configuration, it can be expressed as:
\begin{equation}
\begin{aligned}
    \Gamma_{\text{round}} = &\Gamma_{\text{M}}\Gamma_{\text{FS}}\Gamma_{\text{L}}\Gamma_{\text{G}}\Gamma_{\text{FS}}\Gamma_{\text{L}}\Gamma_{\text{FS}}\Gamma_{\text{M}} \\
                     &\Gamma_{\text{M}}\Gamma_{\text{FS}}\Gamma_{\text{L}}\Gamma_{\text{FS}}\Gamma_{\text{G}}\Gamma_{\text{L}}\Gamma_{\text{FS}}\Gamma_{\text{M}},
\end{aligned}
\end{equation}
The Fox-Li method, a power iteration algorithm, numerically solves for the fundamental mode—the one with the highest transmission efficiency (i.e., largest $|\gamma|$). Starting with an initial field $u_0$, the method iteratively applies the operator:
\begin{equation}
u_{t+1}(x, y) = \Gamma_{\text{round}}\left(u_t(x, y)\right).
\end{equation}
The iteration proceeds until the field distribution converges, which occurs when the change in the transmission factor between consecutive iterations becomes negligible ($|\gamma_t - \gamma_{t-1}| < 10^{-4}$). The resulting field $u_t$ represents the steady-state fundamental mode of the cavity \cite{ref35}.

\subsubsection{Dynamic Cavity Model with MT Motion}
\label{sec:dc}
To analyze the impact of MT motion on the resonant beam, we extend the static model. A transverse velocity of the MT introduces a displacement $[\Delta x, \Delta y]$ during the beam's round-trip time. This movement causes a misalignment between the returning beam and the receiving aperture, which introduces additional intracavity losses and can degrade the link.

We model this effect by modifying the round-trip operator. The misalignment is represented by an aperture mask operator, $\Gamma_{\text{mask}}(\Delta x, \Delta y)$, which nullifies any portion of the returning beam that falls outside the displaced aperture. The modified round-trip propagation operator, $\Gamma'_{\text{round}}$, is then:
\begin{equation}
\Gamma'_{\text{round}} = \Gamma_{\text{mask}}(\Delta x, \Delta y) \cdot \Gamma_{\text{round}}.
\end{equation}
The governing equation for the resonant mode under these dynamic conditions becomes:
\begin{equation}
\Gamma'_{\text{round}} (u) = \gamma' u,
\end{equation}
where $\gamma'$ is the effective transmission factor under motion, the misalignment loss inevitably reduces its magnitude, resulting in $|\gamma'| < |\gamma|$.

The round-trip power efficiency under these dynamic conditions is therefore $|\gamma'|^2$. The reduction of the value increases the intracavity loss. The Fox-Li method can be adapted to find the mode in this dynamic scenario by iteratively applying the modified operator:
\begin{equation}
u_{t+1} = \Gamma'_{\text{round}} (u_t).
\end{equation}
This dynamic model provides a theoretical framework for quantifying the maximum tolerable velocity before beam misalignment leads to link failure, and it confirms that a beam persists under MT motion, making continuous AoA estimation possible.

Therefore, this dynamic model establishes a theoretical framework for quantifying the maximum tolerable speed. The speed limit is numerically determined by iteratively applying the $\Gamma'_{\text{round}}$ operator to solve for the round-trip power efficiency, denoted as $|\gamma'|^2$. The maximum speed threshold is identified as the point where any further speed increase causes $|\gamma'|^2$ to fall below the level compensable by the system gain. Below this threshold, the resonant beam oscillation cannot be sustained.


}

\subsubsection{MUSIC Algorithm for AoA}
With the physical characteristics of the resonant beam established as a stable signal source, we now describe the algorithm used to AoA. In this work, the multiple signal classification (MUSIC) algorithm is employed to estimate the AoA of resonant beams from the light distribution captured by the wavefront sensor. The optical field is spatially sampled by the sensor's $M=1600$ elements ($40 \times 40$ array), yielding a discrete signal vector. For $K$ incident resonant beams, the signal vector $\mathbf{z} \in \mathbb{C}^{M \times 1}$ captured in a single snapshot can be modeled as:
\begin{equation}
\mathbf{z} = \mathbf{A}(\mathbf{\Theta}) \mathbf{s} + \mathbf{n},
\end{equation}
where $\mathbf{A}(\mathbf{\Theta}) \in \mathbb{C}^{M \times K}$ is the array steering matrix, $\mathbf{s} \in \mathbb{C}^{K \times 1}$ is the vector of complex amplitudes for the $K$ source signals, and $\mathbf{n} \in \mathbb{C}^{M \times 1}$ is the additive white Gaussian noise vector with zero mean and covariance $\sigma_n^2 \mathbf{I}$. The steering matrix $\mathbf{A}(\mathbf{\Theta}) = [\mathbf{a}(\mathbf{\Theta}_1), \dots, \mathbf{a}(\mathbf{\Theta}_K)]$ is composed of steering vectors $\mathbf{a}(\mathbf{\Theta}_k)$, each corresponding to a unique arrival direction $\mathbf{\Theta}_k = (\theta_k, \phi_k)$. The $i$-th element of a steering vector is given by:
\begin{equation}
[\mathbf{a}(\theta_k, \phi_k)]_i = \exp\left(j \mathbf{k}_k \cdot \mathbf{r}_i\right),
\end{equation}
where $\mathbf{r}_i$ is the position vector of the $i$-th sensor element. The wave vector $\mathbf{k}_k$ depends on the direction of arrival:
\begin{equation}
\mathbf{k}_k = \frac{2\pi}{\lambda} \begin{bmatrix}
\cos(\phi_k) \cos(\theta_k) & \cos(\phi_k) \sin(\theta_k) & \sin(\phi_k)
\end{bmatrix}^T,
\end{equation}
with $\lambda$ being the wavelength of the resonant beam.

The next step is to analyze the spatial statistics of the received signal by computing its covariance matrix. Given that the resonant beam reaches a steady state, its light field distribution is temporally constant. Therefore, a single snapshot is sufficient to estimate the covariance matrix:
\begin{equation}
\hat{\mathbf{R}}_{z} = \mathbf{z} \mathbf{z}^H,
\end{equation}
where $(\cdot)^H$ denotes the Hermitian transpose. The core of the MUSIC algorithm lies in the eigendecomposition of this covariance matrix, which partitions the vector space $\mathbb{C}^M$ into two orthogonal subspaces: a signal subspace and a noise subspace. The eigendecomposition is:
\begin{equation}
\hat{\mathbf{R}}_{z} = \mathbf{U} \mathbf{\Sigma} \mathbf{U}^H = \begin{bmatrix} \mathbf{U}_S & \mathbf{U}_N \end{bmatrix} \begin{bmatrix} \mathbf{\Sigma}_S & \mathbf{0} \\ \mathbf{0} & \mathbf{\Sigma}_N \end{bmatrix} \begin{bmatrix} \mathbf{U}_S^H \\ \mathbf{U}_N^H \end{bmatrix},
\end{equation}
where the columns of $\mathbf{U}_S \in \mathbb{C}^{M \times K}$ are the eigenvectors corresponding to the $K$ largest eigenvalues and span the signal subspace. The columns of $\mathbf{U}_N \in \mathbb{C}^{M \times (M-K)}$ are the eigenvectors corresponding to the $M-K$ smallest eigenvalues and span the noise subspace.

The fundamental principle of MUSIC is that the steering vectors of the true incident signals are orthogonal to the noise subspace. This orthogonality allows for the estimation of the AoAs by searching for directions whose steering vectors lie outside the noise subspace. This is accomplished by computing the MUSIC pseudo-spectrum:
\begin{equation}
P_{\text{MUSIC}}(\theta, \phi) = \frac{1}{\mathbf{a}^H(\theta, \phi) \mathbf{U}_N \mathbf{U}_N^H \mathbf{a}(\theta, \phi)},
\end{equation}
where $\mathbf{a}(\theta, \phi)$ is the steering vector for a candidate direction $(\theta, \phi)$. When the candidate direction matches a true AoA, its steering vector $\mathbf{a}(\theta, \phi)$ approaches orthogonality with the noise subspace, causing the denominator of the pseudo-spectrum to approach zero. The AoAs are thus identified by finding the $K$ prominent peaks in $P_{\text{MUSIC}}(\theta, \phi)$ over a 2D angular search space.

\subsection{Depth Calculation}
\label{sec:dc}

\begin{figure}[!t]
\centering
\includegraphics[width=3in]{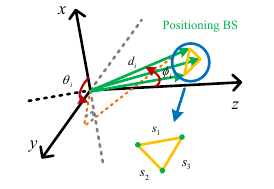}
\caption{Geometric model for BS depth estimation with directional vectors and deployment pattern.}
\label{fig_dp}
\end{figure}
We derive the depths \(d_1\), \(d_2\), and \(d_3\) of three base station given their respective elevation angles \(\phi_{i}\) and azimuth angles \(\theta_{i}\). BSL assumes that the distances between the three base stations (baseline length), denoted as \(s_1\), \(s_2\), and \(s_3\), are known. Base stations deployment pattern as a triangular, as shown in Fig. \ref{fig_dp}.

The direction vectors of the base station, determined by their elevation and azimuth angles, are expressed as:
\begin{equation}
\mathbf{x}_i = \begin{pmatrix}
\cos(\phi_{i})\cos(\theta_{i}) \\
\cos(\phi_{i})\sin(\theta_{i}) \\
\sin(\phi_{i})
\end{pmatrix},
\end{equation}
where \(\mathbf{x}_i\) represents the unit vector in the direction of the base station \(i\) from the MT.

The position of the base station can then be defined in the MT coordinate system by scaling these direction vectors by the respective unknown depths \(d_i\):
\begin{equation}
\mathbf{p}_1 = d_1 \mathbf{x}_1, \quad \mathbf{p}_2 = d_2 \mathbf{x}_2, \quad \mathbf{p}_3 = d_3 \mathbf{x}_3.
\end{equation}

Given the known \(s_1\), \(s_2\), and \(s_3\) between the base station, the following relationships are established:
\begin{equation}
s_1 = \Vert d_2\mathbf{x}_2 - d_3\mathbf{x}_3 \Vert_2,
\end{equation}
\begin{equation}
s_2 = \Vert d_1\mathbf{x}_1 - d_3\mathbf{x}_3 \Vert_2,
\end{equation}
\begin{equation}
s_3 = \Vert d_1\mathbf{x}_1 - d_2\mathbf{x}_2 \Vert_2.
\end{equation}

After analyzing the system, we see that for each base station, we have one unknown depth \(d_i\). We need additional geometric constraints to resolve the exact positions of the base station in three-dimensional space. The mutual \(s_1\), \(s_2\), and \(s_3\) provide exactly these constraints. Specifically, with three base station, we have three depths (\(d_1\), \(d_2\), \(d_3\)) to solve. The (16)-(18) provide the necessary non-linear equations to determine these depths uniquely. Thus, it is the combination of the directional information and the known baseline length that allow us to solve for the depths \(d_1\), \(d_2\), and \(d_3\).

\subsection{Partial Base Station Occlusion Solution}
\label{sec:depth_occlusion}

In real-world scenarios, environmental obstacles often cause the occlusion of certain base stations. To address these challenges, deploying a redundant base station is necessary. This subsection extends the depth computation framework introduced in Section \ref{sec:dc} by exploring base station deployment patterns that can cause partial occlusion.

Consider an equilateral triangular deployment of base stations, where each side of the triangle has a known length \( s \). There are \( N \) base stations uniformly deployed, forming a regular hexagonal grid to ensure scalability and uniform coverage. In an environment with partial occlusion, a subset of these base stations may be obstructed. Suppose \( M \) BSs remain visible within the RB-T's field of view, where \( M > 3 \). This redundancy allows for enhanced robustness and depth estimation through overdetermined systems.

For each visible base station \( i \) (\( i = 1, 2, \dots, M \)), the direction vector is defined by its elevation angle \( \phi_i \) and azimuth angle \( \theta_i \), consistent with Section \ref{sec:dc}. The position vectors in the MT coordinate system are expressed as:
\begin{equation}
\mathbf{p}_i = d_i \mathbf{x}_i ,
\end{equation}
where \( d_i \) is the unknown depth of base station \(i \).

Given the equilateral triangular deployment, the mutual distances between any pairs of base stations are known and uniform, denoted as \( s_{ij} = s \) for all \( i \neq j \). For any pairs of visible base stations \( (i, j) \), the distance constraint is:
\begin{equation}
\Vert \mathbf{p}_i - \mathbf{p}_j \Vert_2 = s.
\end{equation}

Expanding the distance constraint yields:
\begin{equation}
\Vert d_i \mathbf{x}_i - d_j \mathbf{x}_j \Vert_2^2 = s^2,
\end{equation}
Which simplifies to:
\begin{equation}
d_i^2 + d_j^2 - 2 d_i d_j (\mathbf{x}_i^\top \mathbf{x}_j) = s^2,
\end{equation}
where \( \mathbf{x}_i^\top \mathbf{x}_j = \cos(\phi_i)\cos(\phi_j)\cos(\theta_i - \theta_j) + \sin(\phi_i)\sin(\phi_j) \) represents the dot product between the direction vectors of base station \(i \) and base station \(j \).

With \( M \) visible base stations, the number of unique pairwise distance constraints is \( \frac{M(M-1)}{2} \). These constraints form a non-linear system of equations:
\begin{align}
d_1^2 + d_2^2 - 2 d_1 d_2 (\mathbf{x}_1^\top \mathbf{x}_2) &= s^2, \\
d_1^2 + d_3^2 - 2 d_1 d_3 (\mathbf{x}_1^\top \mathbf{x}_3) &= s^2, \\
&\vdots \notag \\
d_{M-1}^2 + d_M^2 - 2 d_{M-1} d_M (\mathbf{x}_{M-1}^\top \mathbf{x}_M) &= s^2.
\end{align}
The system described above is overdetermined since the number of equations \( \frac{M(M-1)}{2} \) exceeds the number of unknowns \( M \) when \( M > 3 \). To solve this system, we adopt a numerical optimization strategy that minimizes the residuals of the distance constraints across all base station pairs. The optimization problem can be formulated as:
\begin{equation}
\min_{\{d_i\}} \sum_{i=1}^{M} \sum_{j=i+1}^{M} \left( d_i^2 + d_j^2 - 2 d_i d_j (\mathbf{x}_i^\top \mathbf{x}_j) - s^2 \right)^2.
\end{equation}

Given the nonlinearity of the equations, iterative methods such as the Levenberg-Marquardt algorithm or gradient descent techniques are employed to find the optimal depths \( \{d_i\} \). The uniformity of the equilateral triangular deployment ensures that the geometric constraints are symmetrically distributed, which aids in achieving stable solutions for the depths \( \{d_i\} \). The redundancy introduced by having \( M > 3 \) visible base stations not only compensates for the occluded ones but also enhances the system's resilience to measurement noise and uncertainties. Specifically, the overdetermined nature of the system allows occlusion and improves the accuracy of the depth estimations.

\section{Self Positioning Method}
This section presents the underlying principles of SL. We begin by describing a general deployment configuration of the RB-T system using a UAV platform and formally defining the SL problem. We then introduce the method for associating base station positions within the MT coordinate system across time instants. Next, we analyze measurement results from a single base station, which leads to the minimal system deployment requirements. Furthermore, we develop an optimization algorithm for scenarios with redundant base stations to enhance positioning accuracy. Finally, we derive the computation method for the output power of the resonant beam and establish evaluation metrics for assessing the performance of both BSL and SL.

\subsection{RBS Integration and Coordinate Transformation}

\begin{figure}[!t]
\centering
\includegraphics[width=2.5in]{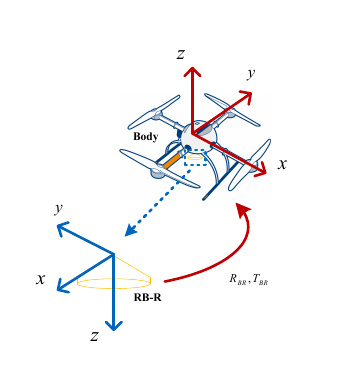}
\caption{Calibration of the DRBP and the UAV.}
\label{fig_2}
\end{figure}
In this subsection, we describe the general installation scheme for RB-T using a UAV as an example, as shown in Fig.~\ref{fig_2}, which illustrates the conventional mounting configuration of the RB-T module. The transformation parameters \(\mathbf{T}_{BR}\) and \(\mathbf{R}_{BR}\) represent the pre-calibrated translation vector and rotation matrix from the MT coordinate system to the UAV body frame. This transformation ensures accurate mapping of position and orientation estimates from the DRBP system into the UAV.

\begin{figure}[!t]
\centering
\subfloat[The MT coordinate system.]{\includegraphics[width=3in]{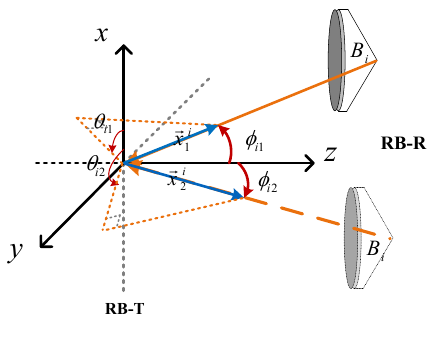}} \\
\subfloat[The world coordinate system.]{\includegraphics[width=3in]{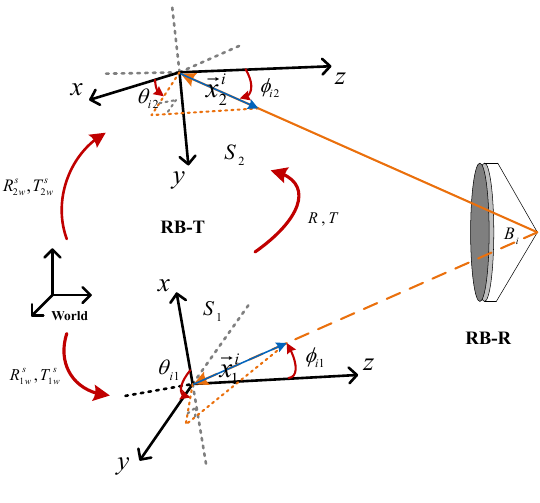}}
\caption{The movement in MT coordinate system and third-view (world) coordinate system. The relative orientations between the MT and base station in the time $t=1$ and $t=2$ are described by angular parameters \(\theta_1, \theta_2, \phi_1, \phi_2\), representing azimuth and elevation angles.}
\label{fig_3}
\end{figure}

Fig.~\ref{fig_3}(a) illustrates the motion of the MT from \(t = 1\) to \(t = 2\) in MT coordinate system, while Fig.~\ref{fig_3}(b) shows the same motion represented in the third-view coordinate system. Consider a network of \(n\) base stations, each uniquely identified as \(B_i\). By combining AoA estimates with the BSL algorithm, the set of 3D coordinates of the base stations in the MT coordinate system at time \(t\) is obtained, denoted as \(P_t = \{\mathbf{p}_1^t, \mathbf{p}_2^t, \dots, \mathbf{p}_n^t\}\). The SL algorithm then estimates the MT's motion by computing the incremental rotation matrix \(\Delta \mathbf{R}\) and displacement vector \(\Delta \mathbf{T}\) between time \(t\) and \(t+1\):
\begin{equation}
[\Delta \mathbf{R}, \Delta \mathbf{T}] = \mathrm{EstimatePose}(P_t, P_{t+1}),
\end{equation}
where \(\mathrm{EstimatePose}(\cdot)\) denotes the SL algorithm that calculates the relative transformation between base station position sets \(P_t\) and \(P_{t+1}\). The computed incremental transformation $[\Delta \mathbf{R}, \Delta \mathbf{T}]$ is then used to recursively update the MT's absolute pose in the initial coordinate system (defined at $t=0$). This process completes the positioning task of the DRBP system between time $t$ and $t+1$.

{\color{blue}
\subsection{Base Station Position Association}
To ensure robust and continuous position tracking, the system performs rigorous data association and manages the base station coordinate sets within the MT coordinate system at time $t$ and $t+1$. Denote the set of base station coordinates in the MT coordinate system at time $t$ as $P_t = \{\mathbf{p}_1^t, \mathbf{p}_2^t, \dots, \mathbf{p}_N^t\}$, and at time $t+1$ as $P_{t+1} = \{\mathbf{p}_1^{t+1}, \mathbf{p}_2^{t+1}, \dots, \mathbf{p}_M^{t+1}\}$. The transformation between consecutive time steps is described by an incremental displacement $\Delta \mathbf{T}$ and an incremental rotation $\Delta \mathbf{R}$.

Given the small displacement and rotational increments between $t$ and $t+1$, a nearest-neighbor search is employed to identify the corresponding point $\mathbf{p}_{\sigma(i)}^{t+1}$ in $P_{t+1}$ for each $\mathbf{p}_i^t \in P_t$:
\begin{equation}
\mathbf{p}_{\sigma(i)}^{t+1} = \arg\min_{\mathbf{p} \in P_{t+1}} \| \mathbf{p} - \mathbf{p}_i^t \|_2,
\end{equation}
where $\|\cdot\|_2$ denotes the Euclidean norm, and $\sigma(i)$ is the index of the matched point in $P_{t+1}$. This yields a set of matched pairs $\{(\mathbf{p}_i^t, \mathbf{p}_{\sigma(i)}^{t+1})\}$, which are used to compute the transformation. A correspondence is considered valid only if the Euclidean distance between the matched points satisfies:
\begin{equation}
\| \mathbf{p}_{\sigma(i)}^{t+1} - \mathbf{p}_i^t \|_2 \leq \delta_{\text{th}},
\end{equation}
where $\delta_{\text{th}}$ is a predefined distance threshold. Points in $P_{t+1}$ that do not satisfy this condition with any point in $P_t$ are identified as new base stations entering the field of view and are initialized for tracking. Since these new stations lack prior positions for motion estimation, they are excluded from the computation of $\Delta \mathbf{T}$ and $\Delta \mathbf{R}$ between $t$ and $t+1$.
Similarly, points in $P_{t}$ that do not have a corresponding point in $P_{t+1}$ within the threshold are considered to have left the field of view and are not used in the transformation computation.
}

\subsection{Minimal Deployment and Constraints}
The minimal deployment of the DRBP system refers to the simplest configuration capable of supporting both SL and BSL. To analyze the requirements for SL, we begin by examining the degrees of freedom involved in estimating the pose change of the MT.

The relative rotation between consecutive MT poses is represented by the rotation matrix $\Delta \mathbf{R}$, which in 3D space can be parameterized using three Euler angles:
\begin{equation}
\Delta \mathbf{R}(\alpha, \beta, \kappa) = \mathbf{R}_x(\alpha) \mathbf{R}_y(\beta) \mathbf{R}_z(\kappa),
\end{equation}
where $\mathbf{R}_x(\alpha)$, $\mathbf{R}_y(\beta)$, and $\mathbf{R}_z(\kappa)$ denote elementary rotations about the x-, y-, and z-axes, respectively. This contributes three rotational degrees of freedom. In addition, the displacement vector $\Delta \mathbf{T}$ introduces three translational degrees of freedom. Thus, the full pose change $[\Delta \mathbf{R}, \Delta \mathbf{T}]$ comprises six DoF in total.

Each base station provides constraints for solving these unknowns. Let $\mathbf{p}_i^t$ and $\mathbf{p}_i^{t+1}$ denote the coordinates of the $i$-th base station in the MT coordinate system at times $t$ and $t+1$, respectively. The following geometric constraint holds:
\begin{equation}
\mathbf{p}_i^{t+1} = \Delta \mathbf{R} \cdot \mathbf{p}_i^t + \Delta \mathbf{T}.
\end{equation}
Each such vector equation provides three constraints. Therefore, two base stations would provide six constraints, which is sufficient to solve for six unknowns.

However, accurately determining the depth of each base station in 3D space requires additional constraints to resolve positional ambiguities. As established in Section~\ref{sec:dc}, the DRBP framework requires at least three base stations to compute depth unambiguously and obtain reliable positional information. Hence, while two base stations are theoretically sufficient for solving the SL problem, a minimum of three are necessary in practice to ensure accurate and unique localization, thereby supporting both SL and BSL functionalities.

\subsection{Positioning method on multiple base station scenarios}
\label{sec:pc}
To enhance the SL accuracy of the DRBP method in scenarios with redundant base stations, we employ an optimization framework to estimate the motion (rotation and translation) of the MT between consecutive time steps. This is achieved by leveraging the corresponding base station positions observed in the MT coordinate system at times $t$ and $t+1$.

Let there be $m$ matched pairs of base station positions between time $t$ and $t+1$. Denote their coordinates as $P_t = \{\mathbf{p}_1^t, \mathbf{p}_2^t, \dots, \mathbf{p}_m^t\}$ and $P_{t+1} = \{\mathbf{p}_1^{t+1}, \mathbf{p}_2^{t+1}, \dots, \mathbf{p}_m^{t+1}\}$ time $t$ and $t+1$ MT coordinate system, respectively. Our goal is to find a rotation matrix $\Delta \mathbf{R}$ and a translation vector $\Delta \mathbf{T}$ such that:
\begin{equation}
\mathbf{p}_i^{t+1} = \Delta \mathbf{R} \cdot \mathbf{p}_i^t + \Delta \mathbf{T}, \quad \text{for } i = 1, 2, \ldots, m.
\label{eqrt}
\end{equation}

The transformation is computed as follows. First, the centroids of the two sets are calculated:
\begin{equation}
\mathbf{p}_c^{t+1} = \frac{1}{m} \sum_{i=1}^{m} \mathbf{p}_i^{t+1}, \quad \mathbf{p}_c^{t} = \frac{1}{m} \sum_{i=1}^{m} \mathbf{p}_i^{t},
\end{equation}
where $\mathbf{p}_c^{t+1}$ and $\mathbf{p}_c^{t}$ are the centroids of $P_{t+1}$ and $P_t$, respectively. Next, the coordinates are centered by subtracting their respective centroids:
\begin{equation}
\mathbf{q}_i^{t+1} = \mathbf{p}_i^{t+1} - \mathbf{p}_c^{t+1}, \quad \mathbf{q}_i^{t} = \mathbf{p}_i^{t} - \mathbf{p}_c^{t}.
\end{equation}

These centered vectors are used to construct a covariance matrix $\mathbf{H}$:
\begin{equation}
\mathbf{H} = \sum_{i=1}^{m} \mathbf{q}_i^{t} (\mathbf{q}_i^{t+1})^\top.
\end{equation}
Subsequently, singular value decomposition (SVD) is applied to $\mathbf{H}$:
\begin{equation}
\mathbf{H} = \mathbf{U} \boldsymbol{\Sigma} \mathbf{V}^\top,
\end{equation}
where $\mathbf{U}$ and $\mathbf{V}$ are $3 \times 3$ orthogonal matrices, and $\boldsymbol{\Sigma}$ is a diagonal matrix of singular values. The optimal rotation matrix is given by:
\begin{equation}
\Delta \mathbf{R} = \mathbf{V} \mathbf{U}^\top.
\label{eq:rotation}
\end{equation}

This rotation minimizes the residual alignment error in the least-squares sense. Finally, the translation vector is computed as:
\begin{equation}
\Delta \mathbf{T} = \mathbf{p}_c^{t+1} - \Delta \mathbf{R} \cdot \mathbf{p}_c^{t}.
\label{eq:translation}
\end{equation}

This transformation optimally aligns the two position sets by mapping the centroid of the rotated initial set to the centroid of the target set. The proposed approach improves SL accuracy in the presence of redundant base stations by robustly estimating MT motion, thereby maintaining positioning accuracy and enhancing system robustness against outlier measurements.

\subsection{Error Analysis Model}

In this section, we analyze the primary sources of error that affect the accuracy of the DRBP system and establish a model to quantify these errors.

\subsubsection{Signal-to-Noise Ratio (SNR)}
The precision of the AoA estimation is fundamentally limited by the SNR of the signal captured by the wavefront sensor. The signal power, $S$, is determined by the output power of the resonant cavity, while the noise power, $N$, arises from various sources within the CMOS sensor.

The received signal power $S$ on the wavefront sensor is given by:
\begin{equation}
S = P_{\text{r}} = \rho P_{\text{out}},
\end{equation}
where $\rho$ is the attenuation factor of the receiver optics, and $P_{\text{out}}$ is the output power of the resonant cavity, which can be modelled as \cite{ref37}:
\begin{equation}
\begin{aligned}
P_{\text{out}} = &\frac{A_b I_s (1 - Rq) \sqrt{V_r}}{1 - Rq V_r + \sqrt{Rq V_r} \left[ 1/(V_r V_s)-V_s \right]} \\
&\cdot \left[ \frac{\eta_{\text{excit}} P_{\text{in}}}{A_g I_s} - \ln \left| \sqrt{Rq V_{s}^{2} V_r} \right| \right],
\end{aligned}
\end{equation}
where $A_{b}$ is the resonant beam area on the gain medium, $A_{g}$ is the cross-sectional area of the gain medium, $I_{s}$ is the saturation intensity of the gain medium, $\eta_{\text{excit}}$ is the excitation efficiency, $Rq$ is the retroreflector reflectivity, and $V_{r}$ is the over-the-air transmission efficiency, equivalent to the squared magnitude of the transmission factor ($|\gamma|^{2}$).

The various noise sources inherent to the CMOS sensor, such as thermal, shot, and flicker noise, can be collectively modelled as additive white Gaussian noise for system-level analysis. This standard model assumes the noise has a flat power spectral density, $N_0$ (in W/Hz), across the system's operational bandwidth, $B$ (in Hz). The total noise power $N$ is therefore the product of the noise density and the bandwidth:
\begin{equation}
    N = N_0 B.
\end{equation}

Consequently, the SNR, which dictates the ultimate performance of the positioning algorithms, can be expressed in dB as:
\begin{equation}
SNR = 10 \log_{10}\left(\frac{S}{N}\right) = 10 \log_{10}\left(\frac{S}{N_0 B}\right) \ \text{dB}.
\end{equation}

\subsubsection{Quantification of Localization Errors}
The performance of the DRBP system is evaluated based on its translational and rotational accuracy. We use the root mean square error (RMSE) as the key metric, as it effectively aggregates estimation errors into a single, comprehensive measure.

For the translational component, the RMSE quantifies the accuracy of the position estimates. Let $N$ be the total number of samples, $\mathbf{t}_{\text{e}, i}$ the estimated position, and $\mathbf{t}_{\text{r}, i}$ the corresponding ground-truth position for the $i$-th sample. The translational RMSE is defined as:
\begin{equation}
RMSE_{\text{t}} = \sqrt{\frac{1}{N} \sum_{i=1}^N \|\mathbf{t}_{\text{e}, i} - \mathbf{t}_{\text{r}, i}\|^2},
\end{equation}
where $\|\cdot\|$ denotes the Euclidean distance. This metric provides a quantitative measure of the overall positioning precision.

For the rotational component, accuracy is equally critical. The rotational RMSE quantifies the error between the estimated and true orientations. Let $\mathbf{R}_{\text{e}, i}$ be the estimated rotation matrix and $\mathbf{R}_{\text{r}, i}$ be the true rotation matrix for the $i$-th sample. The relative rotation error matrix is given by:
\begin{equation}
\mathbf{R}_{\text{diff}, i} = \mathbf{R}_{\text{e}, i}^\top \mathbf{R}_{\text{r}, i}.
\end{equation}
The magnitude of this rotation, representing the angular error, is extracted using the axis-angle formula. The rotational RMSE is then expressed as the root mean square of these angular errors:
\begin{equation}
RMSE_{\text{rot}} = \sqrt{\frac{1}{N} \sum_{i=1}^N \left(\arccos\left(\frac{\text{tr}(\mathbf{R}_{\text{diff}, i}) - 1}{2}\right)\right)^2},
\end{equation}
where $\text{tr}(\cdot)$ is the trace of the matrix. This formula provides a precise measure of the average angular deviation between the estimated and true orientations.

\section{Numerical Analysis}

For the MT-side self-localization application based on DRBP, a fundamental requirement is that the sensing link of the resonant beam remains stable when the MT moves. Therefore, in this section, we first introduce the simulation parameters. Then, we verify the robustness of the system's resonant beam through simulation under MT mobility. After confirming the resonant beam is feasible on the MT mobility scene, we proceed to validate the positioning accuracy of DRBP in a single-MT scenario. Specifically, we will analyze the AoA estimation error of the MUSIC algorithm, discuss the influence of cavity length and baseline on BSL accuracy, and finally investigate the effect of BSL accuracy on the SL accuracy of DRBP. Since the localization process is implemented on the MT side, and MTs share the base stations during localization, we can naturally extend DRBP to any number of MTs once we demonstrate its feasibility and accuracy in a single-MT scenario.

\begin{table}[htbp]
\label{table:2}
\centering
\caption{PARAMETERS FOR RBS SIMULATION \cite{ref19,ref33}}
\begin{tabular}{@{}lll@{}}
\toprule
\textbf{Symbol} & \textbf{Parameter} & \textbf{Value} \\ \midrule
\( r_c \)       & Cat’s eye radius   & 7 mm            \\
\( r_g \)       & Gain medium radius & 7 mm            \\
\( f \)         & Focal length of L1/2 & 50 mm         \\
\( l \)         & Distance between L1/2 and M1/2 & 50.1 mm \\
\( \lambda \)   & resonant beam wavelength & 1064 nm       \\
\( Rq \)         & Reflectivity of M1/2 & 0.99            \\
\( \eta_{excit} \) & Excitation efficiency & 0.72       \\
\( V_s \)       & Loss factor in medium & 0.99           \\
\( I_s \)       & Medium saturated intensity & \( 1.26 \times 10^6 \) W / m$^2$ \\
\( \rho \)      & Attenuation factor & \( 10^{-4} \)    \\
\( N_s \)       & wavefront sensor pixel number & 16384              \\
\( S_N \)       & FFT sampling number & 16384            \\
\( G \)         & Computation window expand factor & 3    \\ \bottomrule
\end{tabular}
\end{table}
\subsection{Parameters}
\begin{figure*}[!t]
\centering
\subfloat[Transmission factor vs. time]{\includegraphics[width=2.5in]{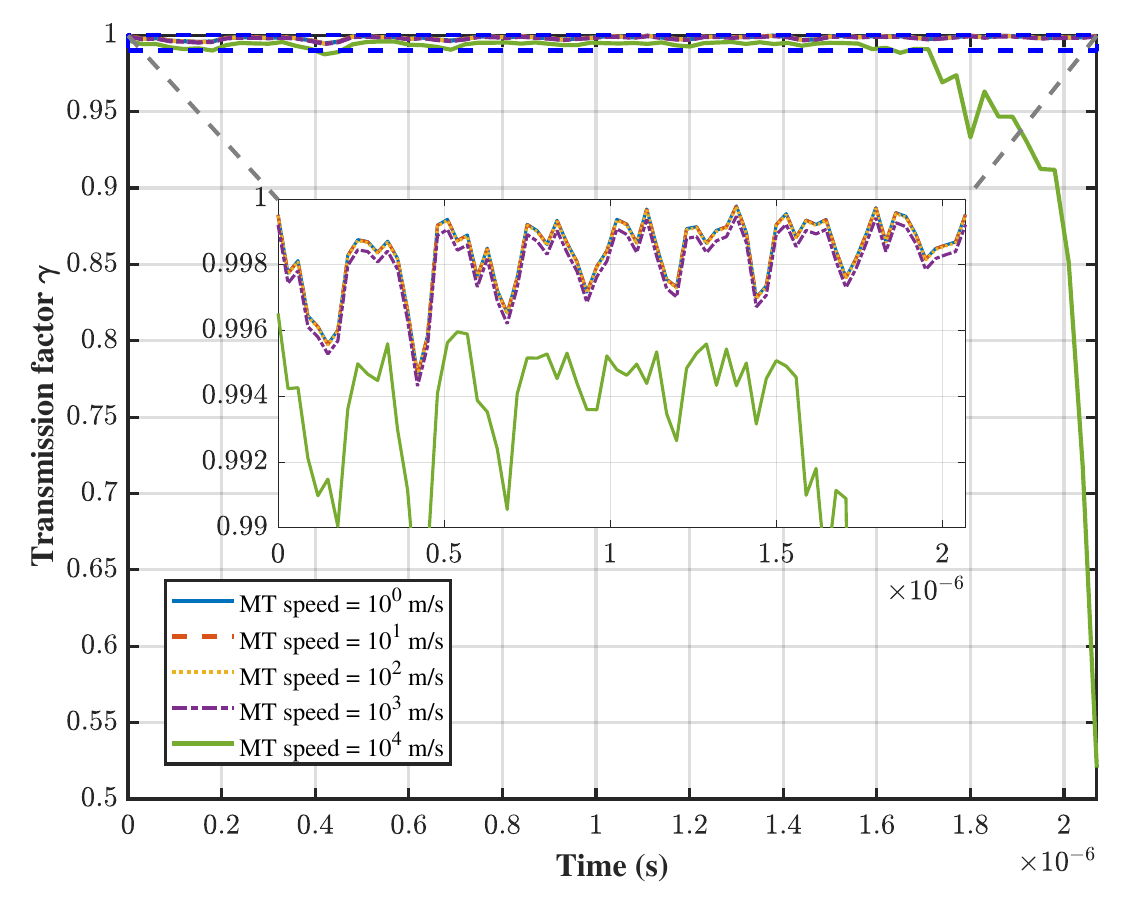}} \ \ \ 
\subfloat[Intracavity power vs. time]{\includegraphics[width=2.5in]{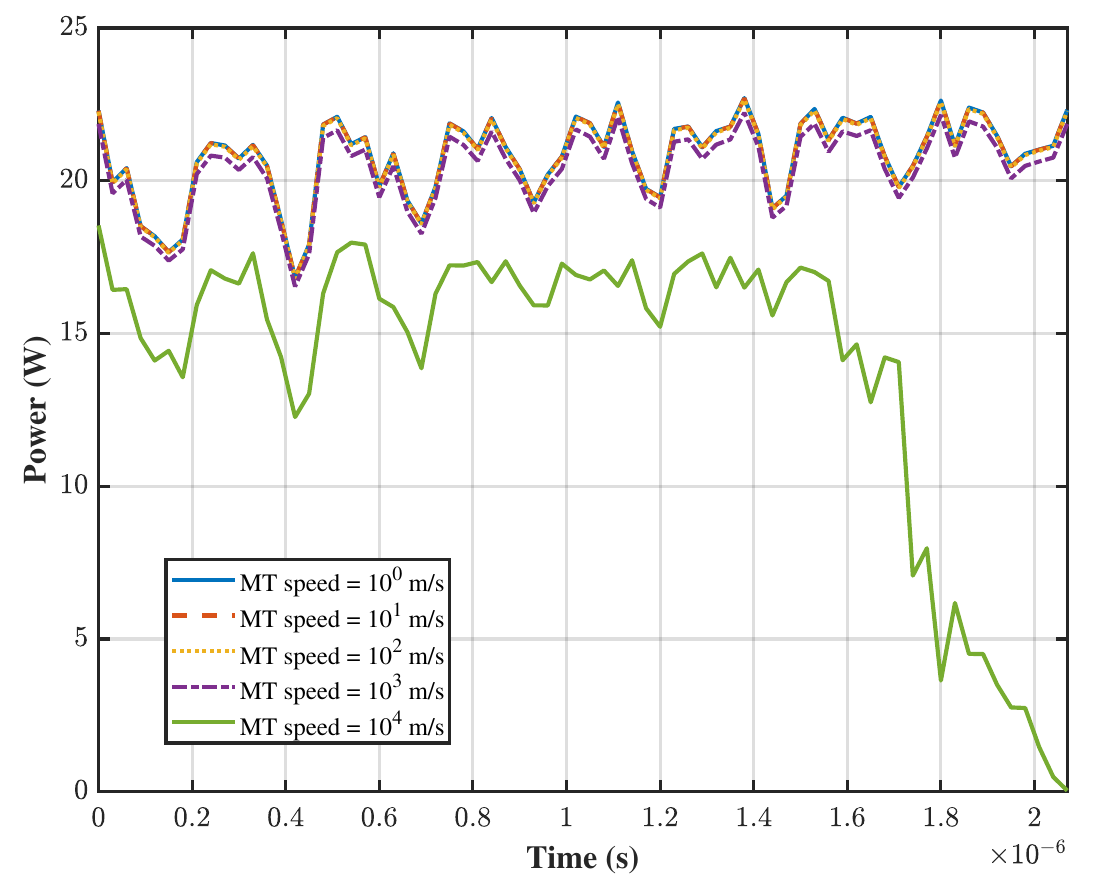}}
\caption{{\color{blue}Time-domain variation of transmission factor and intracavity power under different MT speeds. (a) Time-domain variation of the transmission factor. (b) Corresponding time-domain variation of the intracavity power under 100 W pump power and a 7 mm aperture diameter.}} \label{fig_dynamic_sim}
\end{figure*}

This section outlines the parameter settings made for the numerical simulations of the DRBP. The RBP can sense and achieve millimetre-level positioning accuracy at $2$ m through experimental verification. Additionally, simulations confirmed that a resonant beam could achieve emission and energy transfer at $10$ m \cite{ref19}, and the TIM system can change the light phase and propagation direction based on reference \cite{ref32}. Therefore, our simulation parameters are based on reference \cite{ref19,ref33}, as shown in Table~II.

Key component parameters, such as those for the gain medium, were based on a standard diode-pumped Nd laser system known for its effectiveness in such simulations. The Fox-Li iterative algorithm was utilized to model the resonant beam self-reproducing modes accurately. In line with the recommendations in \cite{ref34}, careful sampling and zero-padding were applied during fast fourier transform (FFT) calculations to prevent aliasing, ensuring the sampling precision necessary for accurate resonator mode analysis.

{\color{blue}
\subsection{Analysis of Dynamic Mobility Performance}

The operational principle of the DRBP system is based on a resonant beam for position. Therefore, sustaining a resonant beam during the movement of the MT is a fundamental requirement. In this subsection, we evaluate the sustainability of resonant beam by analyzing the intra-cavity power of the DRBP system at various MT velocities. The analysis is based on the dynamic resonant cavity model established in Section~\ref{sec:dc}.

The changes of transfer factor and intra-cavity power with different movement velocities are depicted in Fig. 7. Fig.~\ref{fig_dynamic_sim}(a) shows that the system’s transmission factor stably maintains around 0.998 as the MT speed increases from 1 m/s to 1000 m/s. Correspondingly, Fig.~\ref{fig_dynamic_sim}(b) shows that the intra-cavity power remains at a high level of approximately 21 W (with $P_{in}=100W$). This indicates that the motion-induced dynamic loss is low within this speed range. The gain medium stably compensates for the loss, and the system consistently maintains the resonant beam. Only in the extreme case, if the speed reaches 10$^4$ m/s (approximately Mach 29), does the accumulated beam misalignment sharply increase the loss and subsequently interrupt the link.

This result confirms the dynamic feasibility of the DRBP system, demonstrating that the resonant beam maintains stable link connectivity under any practical MT speed. Based on this dynamic stability, we recursively execute the BSL and SL processes using discrete sensing snapshots of the resonant beam information. Therefore, with the dynamic link feasibility confirmed, we now proceed to analyze the corresponding positioning accuracy of the DRBP algorithm.
}

\subsection{Analysis of AoA Accuracy}
\label{sec:a_aoa}

\begin{figure}[!t]
\centering
\subfloat[AoA error with $\phi = 4^{\circ}$]{\includegraphics[width=2.7in]{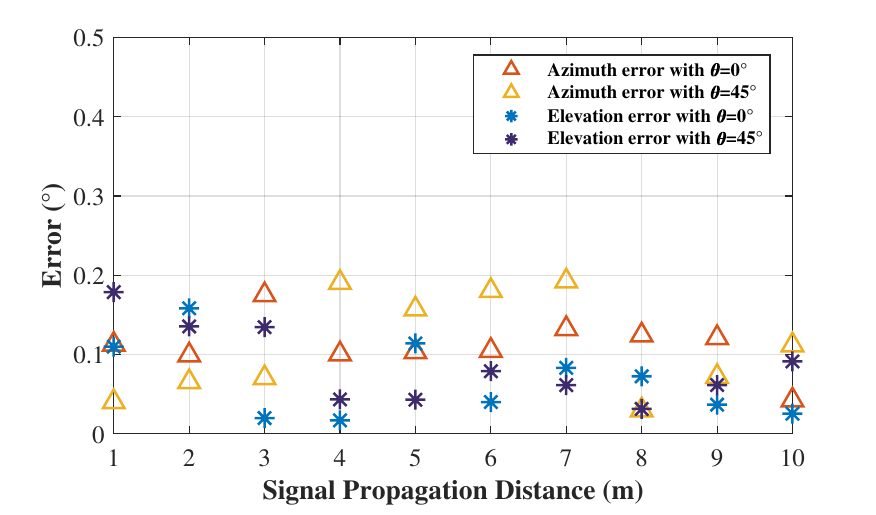}} \\
\subfloat[AoA error with $d=8$ m]{\includegraphics[width=2.7in]{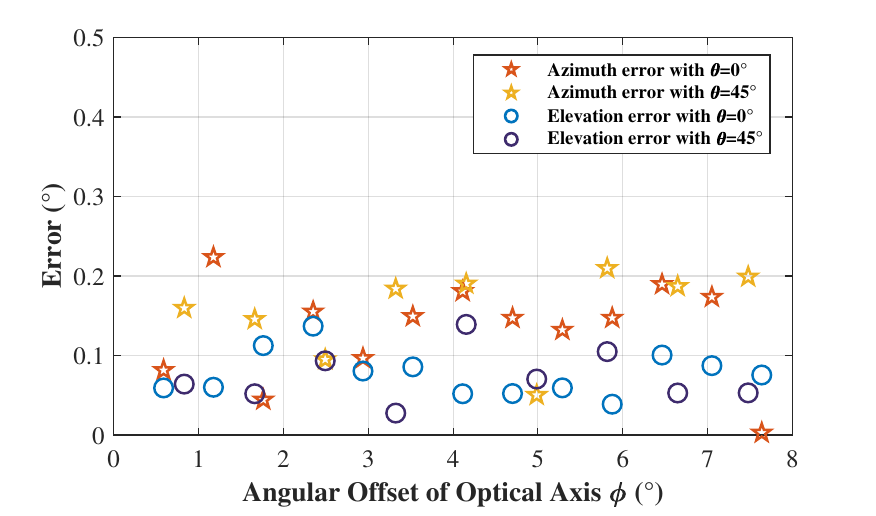}}
\caption{Analysis of AoA errors: (a) Estimation error variation for azimuth and elevation at $\theta=0^{\circ}$ and $\theta=45^{\circ}$ over different signal propagation distances. (b) Azimuth and elevation errors at a fixed azimuth angle were analyzed across varying elevation angles at a signal propagation distance of $8\ m$.}
\label{fig_ana_aoa}
\end{figure}

\begin{figure}[!t]
\centering
\subfloat{\includegraphics[width=2.7in]{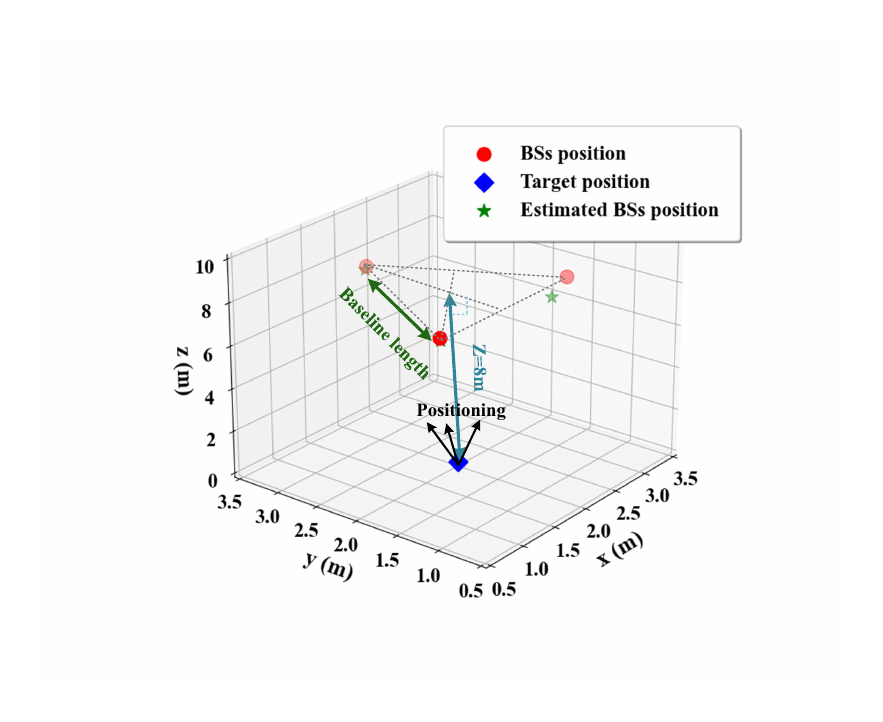}}
\caption{Illustration of the BSL position estimation process under AoA errors uniformly distributed from $0^{\circ}$ to $0.2^{\circ}$. ($s_{bl}=2$ m and $Z = 8$ m)}
\label{fig_61}
\end{figure}

\begin{figure}[!t]
\centering
\subfloat{\includegraphics[width=2.7in]{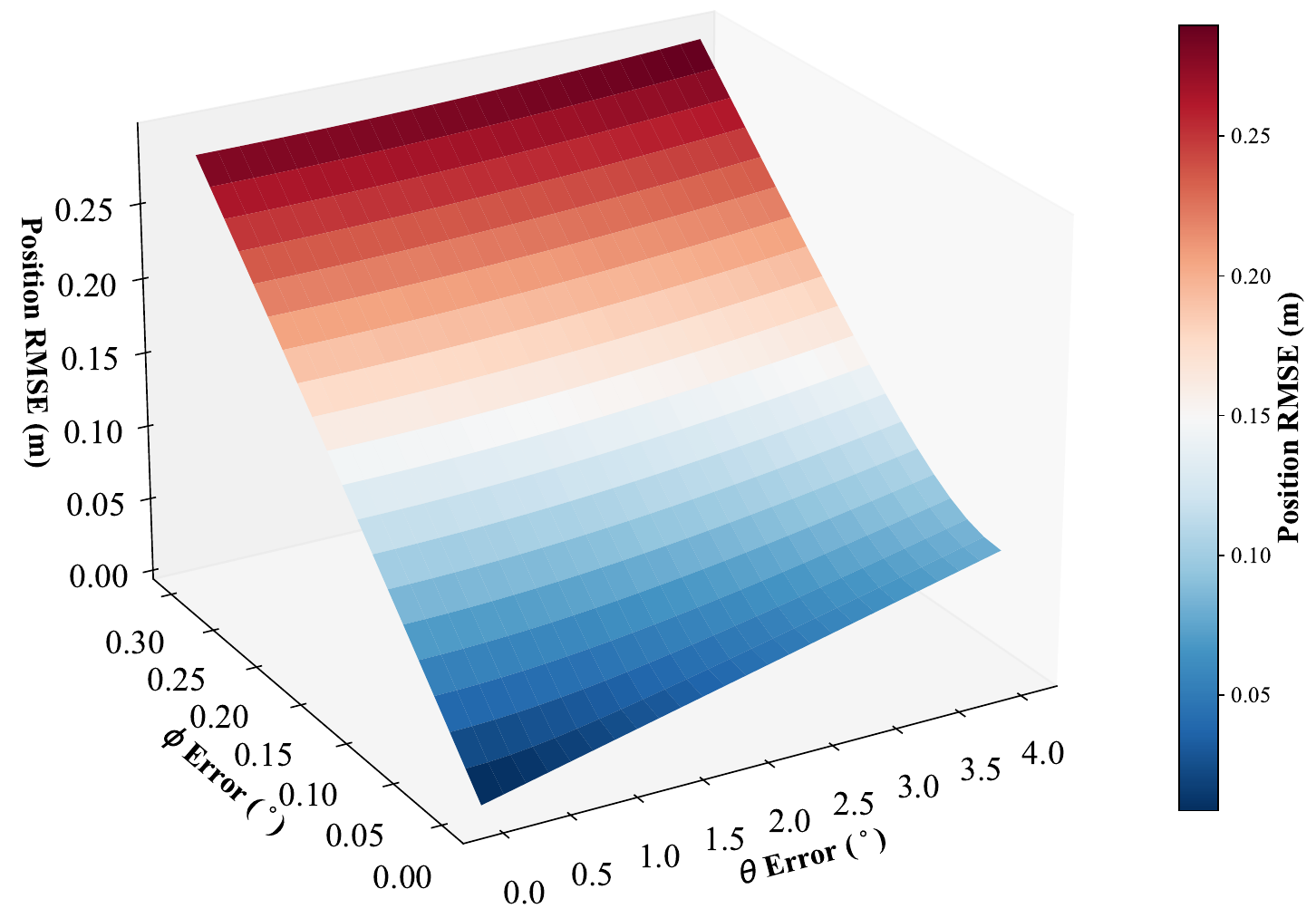}}
\caption{Impact of AoA errors on SL positioning accuracy for $s_{bl} = 2$ m and $Z = 8$ m.}
\label{fig_62}
\end{figure}

We used the AoA estimation algorithm demonstrated in Section~\ref{sec:ac} to calculate the incident angle of the resonant beam and compared it with the actual incident angle. This comparison verified the high accuracy of the proposed AoA estimation method.

In Fig.~\ref{fig_ana_aoa}, we depict the distribution of AoA estimation errors for different resonant beam cavity lengths and various elevation angles, where the pump power \(P_{\text{in}}\) is set to 150 W. Here, $\phi$ represents the elevation angle. An increase in $\phi$ indicates a greater off-axis deviation between the MT and the base station, which in turn increases the beam's propagation distance. This deviation increases the propagation distance of the corresponding resonant beam, which may affect the RBS signal transmission power and the AoA accuracy.

However, as shown in Fig.~\ref{fig_ana_aoa}, the AoA errors are uniformly distributed between $0^{\circ}$ and $0.2^{\circ}$ as the cavity length and \(\phi\) increase. This indicates that AoA accuracy is not significantly affected by changes in cavity length and \(\phi\). This result is attributed to the strong signal power of RBS as an open-cavity laser, where the signal power variation within the FoV is small \cite{ref30}, leading to a uniform distribution of AoA estimation errors within the FoV.

\subsection{Analysis of BSL Accuracy}
\label{sec:ala}
\begin{figure}[!t]
\centering
\subfloat{\includegraphics[width=2.7in]{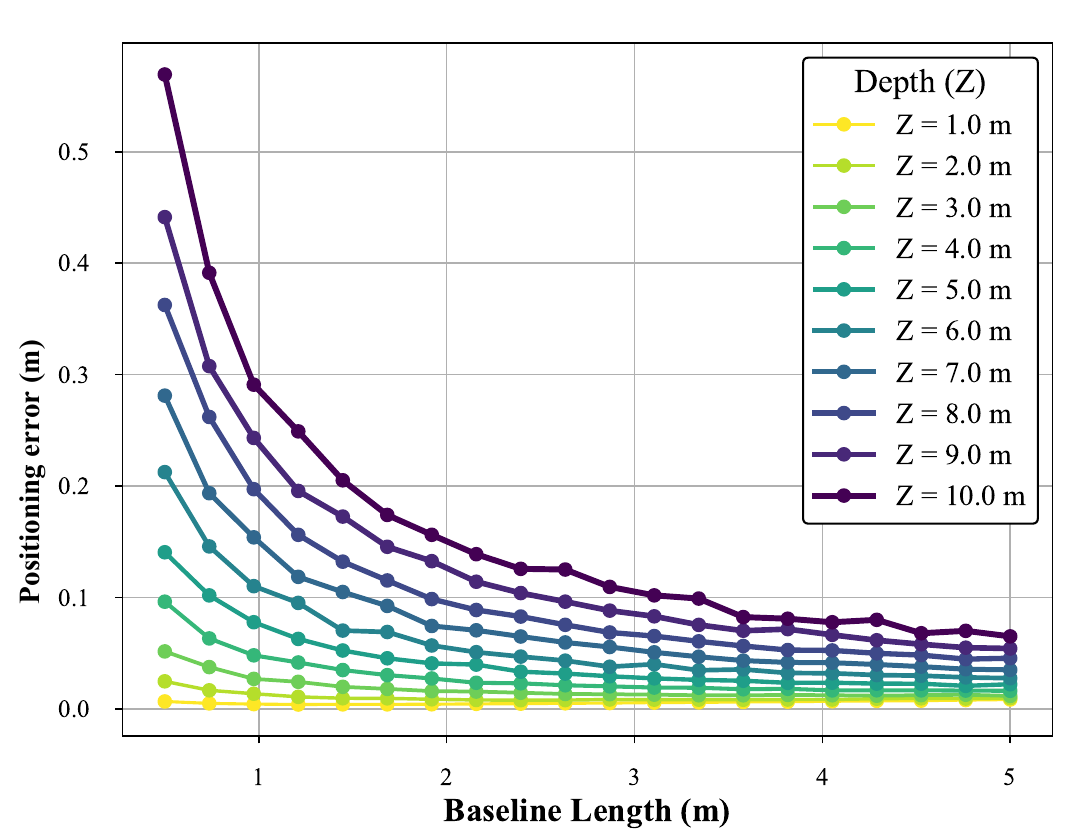}}
\caption{The RMSE of BSL positioning error as the $s_{bl}$ increases, with the plane of $Z\in[1,10]$.}
\label{fig_63}
\end{figure}

\begin{figure}[!t]
\centering
\subfloat{\includegraphics[width=2.7in]{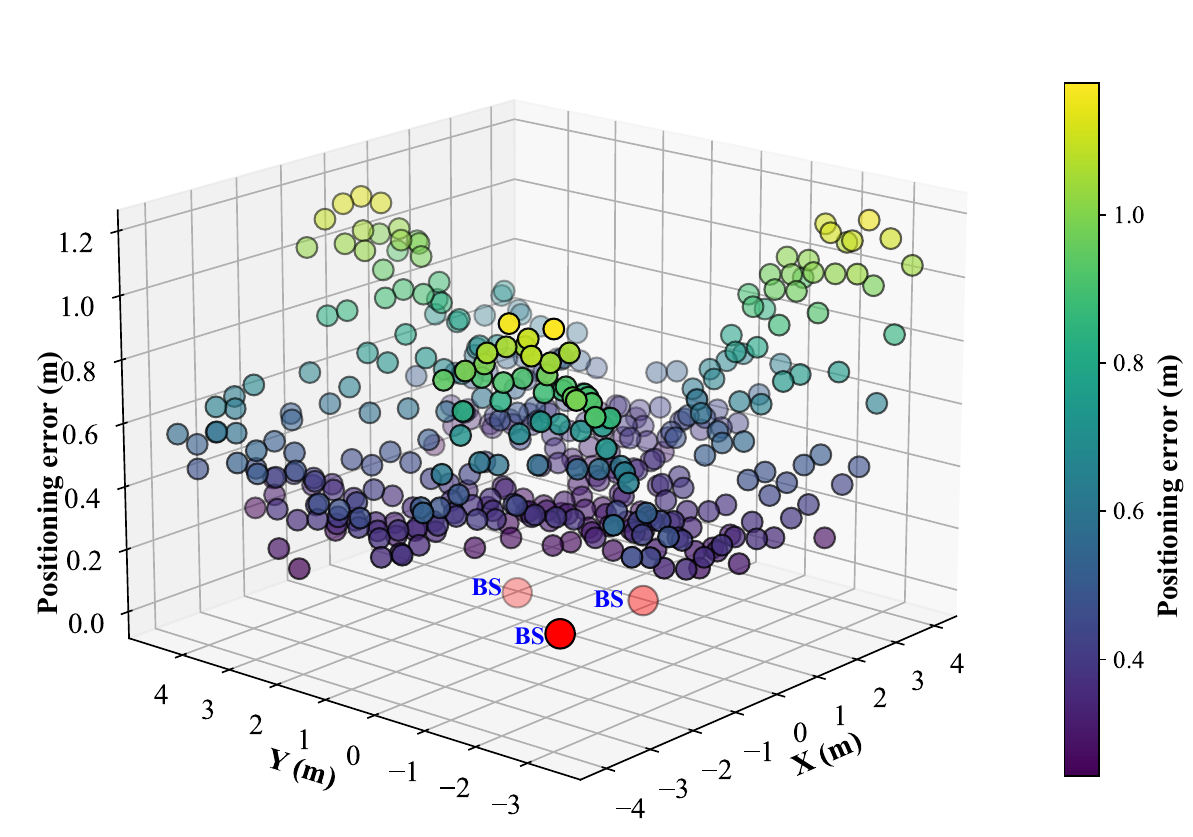}}
\caption{The RMSE distribution of BSL positioning errors for different MT projection positions on the base stations plane with $s_{bl}=2$ m, $Z=8$ m and AoA errors uniformly distributed from $0^{\circ}$ to $0.2^{\circ}$}
\label{fig_64}
\end{figure}

The BSL is crucial for implementing DRBP. Accurate depth calculation is essential for achieving base station positioning. As discussed in Section~\ref{sec:pe}, the BSL positioning accuracy is primarily influenced by the baseline length and AoA errors. In this section, we examine the impact of variations in these parameters on the positioning error. The base stations are uniformly deployed with \(\text{Baseline length} = s_1 = s_2 = s_3 = s_{\text{bl}}\). To reduce the randomness of positioning accuracy, we repeat the calculation 50 times and use the RMSE for error analysis.

Fig.~\ref{fig_61} is a schematic of the BSL process. Under the conditions of an $AoA\ error = 0.2^{\circ}$ and $s_{bl} = 2\ m$, the positioning RMSE is $0.18$ m, demonstrating the effectiveness of BSL. The positioning accuracy is shown in Fig.~\ref{fig_62}, demonstrating the impact of AoA error on BSL positioning accuracy. As the AoA error increases, the positioning error also increases, with elevation angle error having a more pronounced impact on the positioning results. Furthermore, as the elevation angle error grows, the influence of the azimuth angle on the positioning results diminishes. Consequently, the positioning error surface illustrated in Fig.~\ref{fig_62} takes on a saddle-shaped form.

As derived in Section~\ref{sec:a_aoa}, the AoA errors are uniformly distributed in $[0,0.2]^{\circ}$. Therefore, when the AoA error distribution is fixed, we illustrate the impact of baseline length and $\textbf{Z}$ on positioning accuracy, as shown in Fig.~\ref{fig_63}.
It can be observed that the RMSE of positioning error significantly increases as $\textbf{Z}$ grows. This is attributed to the amplification of AoA error effects over longer distances. However, the error value is significantly reduced with the increase in baseline length. The BSL can maintain sub-decimeter-level accuracy at a baseline length of $2$ m, even when the MT is at considerable depths.

Fig.~\ref{fig_64} depicts the positioning error results for an MT located at different positions at $Z=8$ m plane. It can be observed that the positioning error is smallest within the base stations region, achieving centimetre-level accuracy. As the distance from the base stations region increases, the error also increases. Therefore, increasing the baseline length during deployment is recommended to expand the enclosed area of the base stations, which will provide higher BSL positioning accuracy and improve the precision of the SL positioning in the DRBP system.

\subsection{Analysis of SL Accuracy}
\begin{figure}[!t]
\centering
\includegraphics[width=2.7in]{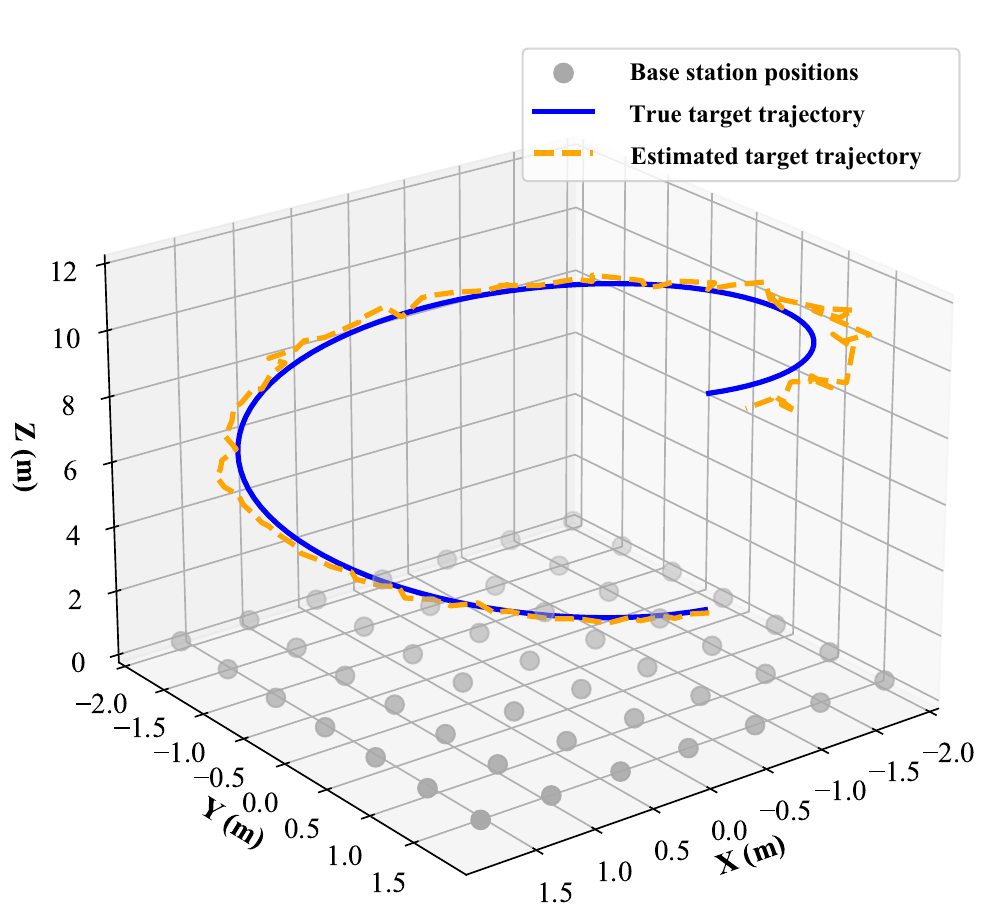}
\centering
\caption{Base station positions and MT trajectory: true and estimated MT trajectories with base station positioned in a 7x7 grid, spaced 0.6 m apart.}
\label{fig_71}
\end{figure}

\begin{figure}[!t]
\centering
\includegraphics[width=2.7in]{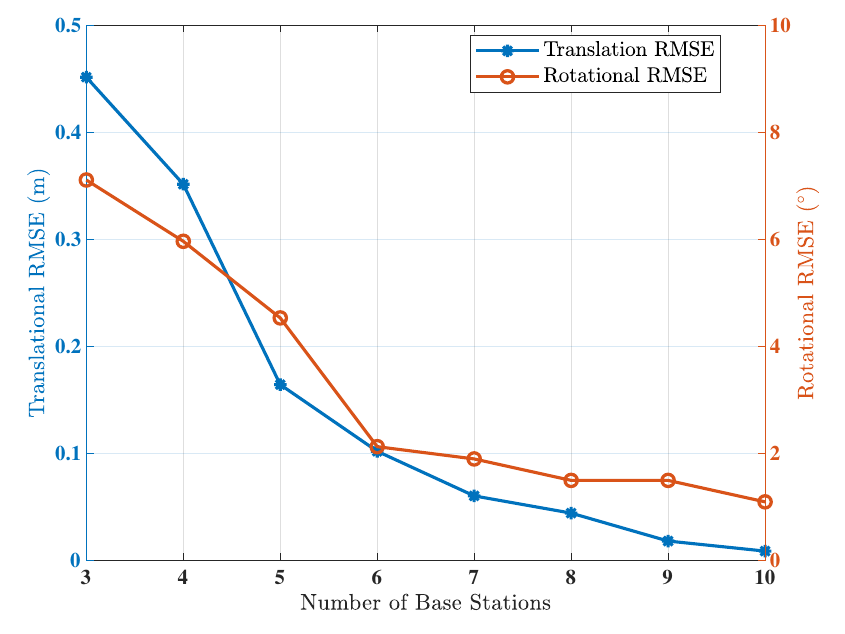}
\centering
\caption{Base station count impact on SL positioning accuracy: translation and rotation error analysis as the number of base station varies, with angle error uniformly distributed from $0^{\circ}$ to $0.2^{\circ}$.}
\label{fig_72}
\end{figure}
In this section, based on the previous analysis of AoA accuracy and BSL positioning accuracy, we evaluate the trajectories obtained by SL to verify the DRBP system’s effectiveness.

Fig.~\ref{fig_71} shows the base station positions and MT trajectory, where the true and estimated MT trajectories are plotted. The base stations are positioned in a 7x7 grid with a spacing of $0.6$ m. During the target operation, the DRBP tracked 6 base stations while the MT completed a full rotation around the z-axis. The true MT trajectory is depicted in blue, and the estimated MT position is shown in orange with the $RMSE_{\text{t}}=0.08$ m and $RMSE_{\text{rot}} = 2^{\circ}$. This result indicates that the DRBP system can effectively estimate the position of the MT in the initial coordinate system at any given moment, while ensuring centimetre-level accuracy.

{\color{blue}To characterize the system's robustness and understand its operational limits, we perform an ablation analysis on the self-localization accuracy as a function of the number of visible base stations, with the results presented in Fig.~\ref{fig_72}.} The translation RMSE is presented on the left y-axis, and the rotational RMSE is presented on the right y-axis. The AoA error is uniformly distributed between $[0,0.2]^{\circ}$ and the baseline length is $1$ m. As the number of base stations increases from 3 to 10, the translation RMSE drops from approximately $0.4$ m to below $0.05$ m, and the rotational RMSE decreases from around $8^{\circ}$ to less than $1^{\circ}$. These results indicate that a higher density of base stations enhances the DRBP precision.

Combining the BSL error analysis from Section~\ref{sec:ala}, SL analyse results highlight the balance between base station deployment pattern and SL accuracy. While a denser base station grid helps reduce trajectory errors, excessive density can shorten the baseline length, thereby reducing positioning accuracy and ultimately affecting the SL results. Consequently, in practical deployments, we must carefully consider baseline length and base station density to optimize SL accuracy, especially in complex environments.

{\color{blue}Finally, we compare the proposed DRBP system with the traditional RBP framework in Table III. The comparison is structured along two primary axes: key performance metrics (translational and rotational accuracy) and fundamental architectural characteristics (scalability and concurrency). The results show that the self-localization accuracy of the DRBP system is slightly lower than that of traditional RBP. This primarily results from the error in the initial BSL phase, which propagates to the subsequent SL process. In contrast, the traditional RBP system relies on pre-calibrated base stations with known positions, thereby avoiding this type of error propagation and achieving higher accuracy.

However, the DRBP system achieves rotation estimation capability within its self-localization architecture. Furthermore, it dynamically localizes passive base stations during operation. This eliminates the dependence on fixed infrastructure and provides the potential for real-time coverage expansion. Additionally, the MT-centric localization mechanism inherently grants the system high concurrency performance. In summary, the DRBP system trades a slight decrease in translational accuracy for a significant enhancement in concurrency support and system scalability.}

\begin{table}[htbp]
\label{table:3}
\centering
\caption{{\color{blue}Comparison of Self-Localization Performance} \cite{ref19,ref33}}
\begin{tabular}{@{}lll@{}}
\toprule
\textbf{}              & \textbf{Traditional RBP}   & \textbf{DRBP (Proposed)} \\ \midrule
Translational (m) & \textbf{0.02}              & 0.08 \\
Rotational (deg)  & N/A                        & \textbf{2.0} \\
Scalability       & Low & \textbf{High (Dynamic Extension)} \\
Concurrency       & Low & \textbf{High (Central Free)} \\ \bottomrule
\end{tabular}
\end{table}

\section{Conclusion}

To enhance the extensibility of RBP, we introduce the implementation of DRBP, which can simultaneously self-localization and base station localization. We propose the BSL method and the SL method. The BSL method is capable of base station localization. The SL method enables self-localization based on base stations‘ positions. The integration of the BSL method with the SL method enables the MT to perform base station localization and self-localization during movement continuously. Our numerical analysis demonstrated that DRBP maintains a sub-decimeter positioning accuracy support for MT. The BSL has an RMSE of \(0.18\) m at a baseline length of $2$ m and a depth of $8$ m, even when facing AoA errors up to \(0.2^{\circ}\). These results also indicated that BSL positioning errors are minimized when the MT is within the area enclosed by the base station, highlighting the importance of base station deployment. Based on this accuracy, the SL achieved RMSE$_{t}=0.08$ m and RMSE$_{rot} = 2^{\circ}$ using a 7x7 grid of base stations with $0.6$ m spacing. Moreover, we proved that SL’s accuracy improved with increased base stations.

In conclusion, the DRBP can simultaneously self-localization and base station localization, enabling the extension of the base station quantity. Future work will focus on enhancing the integration of DRBP with other sensors and improving its stability in specific occlusion scenarios. DRBP system offers scalable positioning services, allows the passive deployment of base stations, focuses energy without signal scattering, and eliminates the need for beamforming control.

\end{document}